\documentclass{elsart}

\begin{document}

\begin{frontmatter}
\title{An Efficient Molecular Dynamics Scheme for Predicting
Dopant Implant Profiles in Semiconductors}

\author{Keith M. Beardmore} and
\author{Niels Gr{\o}nbech-Jensen}
\address{Theoretical Division, Los Alamos National Laboratory,\\
Los Alamos, NM 87545, USA}

\begin{abstract}
   We present a highly efficient molecular dynamics scheme for
calculating the concentration profile of dopants implanted in
group-IV alloy, and III-V zinc blende structure materials.
Our program incorporates methods for reducing computational
overhead, plus a rare event algorithm to give statistical accuracy
over several orders of magnitude change in the dopant concentration.
  The code uses a molecular dynamics
(MD) model, instead of the binary collision approximation (BCA) used
in implant simulators such as TRIM and Marlowe, to describe ion-target
interactions. Atomic interactions are described by a combination of
`many-body' and screened Coulomb potentials. Inelastic
energy loss is accounted for using a Firsov model, and electronic
stopping is described by a Brandt-Kitagawa model which
contains the single adjustable parameter for the entire scheme.
Thus, the program is easily extensible to new ion-target
combinations with the minimum of tuning, and is predictive over a
wide range of implant energies and angles.
   The scheme is especially suited for calculating profiles due to low
energy, large angle implants, and for situations where a predictive
capability is required with the minimum of experimental validation.
We give examples of using our code to calculate concentration
profiles and 2D `point response' profiles of dopants in crystalline
silicon, silicon-germanium blends, and gallium-arsenide.
We can predict the experimental profile over five orders of magnitude
for $\langle$100$\rangle$ and $\langle$110$\rangle$
channeling and for non-channeling implants at
energies up to hundreds of keV.

\end{abstract}

\begin{keyword}
Molecular Dynamics, Ion Implant, Binary Collision, SIMS, Semiconductors.
\end{keyword}

\end{frontmatter}

\date{15 September 1998}

\section{Introduction}
The continuing quest for greater processor performance demands
ever smaller device sizes.
The effort to realize these ultra shallow junction devices has resulted
in current industry trends, such as the use of low energy, high mass,
high dose, and large angle implants, to create abrupt dopant profiles.
The experimental measurement of such profiles
is challenging, as effects that are negligible
at high implant energies become increasingly
important as the implant energy is lowered. For example, 
the measurement of dopant profiles by secondary ion mass
spectrometry (SIMS) is problematic for very low energy
(less than 10 keV) implants, due to limited depth resolution of
measured profiles.
Also, refining SIMS protocols to obtain profiles for new ion-target
combinations, e.g. In, Sb, or N implants or Si$_{1-x}$Ge$_x$ targets
is not a trivial problem.

The use of computer simulation as an alternative method to
determine dopant profiles is well established.
Binary collision approximation (BCA)
codes have traditionally been used, however
such simulations become unreliable at low ion energies\cite{rob74,eck91}.
The BCA approach breaks down when
multiple collisions
or collisions between moving atoms
become significant, or
when the crystal binding energy is of the same order as the energy
of the ion.
Such problems are clearly evident when one attempts to use the BCA to
simulate channeling in semiconductors; here the interactions between
the ion and target are neither binary nor collisional in nature,
rather they occur as many simultaneous soft interactions which steer
the ion down the channel. 

A more accurate, alternative to the BCA, is the use of molecular
dynamics (MD) simulation to calculate ion
trajectories\cite{har88,cai96}.
However, the computational cost of traditional MD simulations precludes
the calculation of the thousands of ion trajectories necessary to
construct a dopant profile.
   Here we present a highly efficient MD-based scheme, that is optimized
to calculate the concentration profiles of ions
implanted into semiconductors.
The algorithms are incorporated into our implant modeling
molecular dynamics code\cite{bea98}, REED-MD\footnote{Named for
`Rare Event Enhanced Domain following
Molecular Dynamics'.}.
Our program has previously been demonstrated to describe the
low dose (zero damage) implant of
As, B, and P ions with energies in the sub MeV range
into crystalline Si in
$\langle$100$\rangle$, $\langle$110$\rangle$, and
non-channeling directions, and also into amorphous
Si\cite{cai96,bea98,cai98}.
We have now extended our model to any ion
species, and to other diamond crystal substrates, such as
C, Ge, SiC, Si$_{1-x}$Ge$_x$, and GaAs.
A model for ion induced damage has also been added to the program
so that high dose implants can be simulated.

\section{Molecular Dynamics Model}
The basis of the molecular dynamics model is a collection of empirical
potential functions that describe interactions between atoms and give
rise to forces between them. In addition to the classical interactions
described by the potential functions,
the interaction of the ion with the
target electrons is required for ion implant simulations,
as this is the principle way in which the ion loses energy.
Another necessary ingredient is a description of the target material
structure, including thermal vibration amplitudes.

   Interactions between target atoms are modeled by derivatives
of the many-body potential
developed by Tersoff\cite{ter88,smi92}.
   ZBL `pair specific' screened Coulomb potentials\cite{zei85,mor97}
are used to model interactions for common ion-target combinations.
For other combinations, the ZBL `universal' potential is used.
The `universal' potential is also used to describe
the close-range repulsive part of the Tersoff potentials.

We include energy loss
due to inelastic collisions, and energy loss due to
electronic stopping as two distinct mechanisms.
The Firsov model\cite{fir59} is used to describe the loss of kinetic
energy from the ion due
to momentum transfer between the electrons of the ion and target atom.
We implement this using a velocity dependent pair potential, as derived
by Kishinevskii\cite{kis62}.

A modified Brandt-Kitagawa\cite{bra82} model,
that involves both global and local contributions to the
electronic stopping is used for electronic energy
loss\cite{cai96,cai98}.
This model contains the single fitted parameter in our scheme,
$r^0_s$, the `average' one electron
radius of the target material experienced by the ion.
This is adjusted to account for oscillations in the $Z_1$ dependence
of the electronic stopping cross-section.
The parameter is fit once for each ion-target combination and is then
valid for all ion energies and incident directions.
By using a realistic stopping model,
with the minimum of fitted parameters,
we obtain a greater transferability to the modeling of implants
outside the fitting set.

   In the calculations presented here, the target is a \{100\}
diamond crystal with a surface oxide layer.
  The oxide structure was obtained from annealing a periodic
SiO$_2$ sample with the density constrained to that
estimated for grown surface oxide\cite{mot98}.
   Thermal vibrations of atoms are modeled by displacing atoms
from their lattice sites using a Debye model.
  For high dose implants, the accumulation of damage within
the target is described by a simple Kinchin-Pease\cite{kin55}
model.
Target properties are either species dependent,
e.g., local electron density,
or are obtained by interpolation from known values for single
element materials, e.g., lattice constant and Debye temperature.

\section{Efficient Molecular Dynamics Algorithms}
We apply a combination of methods to increase the efficiency of
this specific type of simulation.
   Infrequently updated neighbor lists\cite{bea95,ver67}
are employed to minimize the time spent in force calculations.
  The paths of the atoms are integrated using Verlet's
algorithm\cite{ver67}, with a variable timestep that
is dependent upon both
kinetic and potential energy of atoms\cite{bea95}.

It is infeasible to
calculate dopant profiles by full MD simulation, as
computational requirements
scale approximately as $u^{4}$, where $u$ is the initial ion velocity.
We have developed a modified MD scheme
which is capable of producing accurate dopant profiles with a much
smaller computational overhead.
We continually create and destroy target atoms, to follow the domain of
the substrate that contains the ion. Material is built in front
of the ion, and destroyed in its wake.
Hence, the ion experiences the equivalent of a complete crystal,
but the cost of the algorithm is only O($u$).

To further improve efficiency, we use three other approximations.
The moving atom approximation\cite{har88} is used to
reduce the number of force calculations.
Atoms are divided
into two sets; those that are `on' have their positions integrated,
and those that are `off' are stationary.
At the start of the simulation, only the
ion is turned on.
Some of the `off' atoms will be used in the force calculations
and will have forces assigned to them. If the resultant force
exceeds a certain threshold, the atom is turned on.
   We use two thresholds in our simulation; all atoms interacting
directly with the ion are turned on immediately (zero threshold),
and other atoms are
turned on if the force exceeds 8.0$\times$10$^{-9}$~N.

For high ion velocities, we do not need to use a many-body potential
to maintain the stable diamond lattice;
a pair potential is sufficient, as only
repulsive interactions are significant.
Hence, at a certain ion velocity we switch from the complete many-body
potential to a pair potential approximation for the target atom
interactions.
We make a further approximation for still higher ion energies,
where only
ion-target interactions are significant in determining the ion path.
This approximation,
termed the recoil interaction approximation\cite{nor95},
brings the MD scheme close to many BCA implementations.
The major difference between
the two approaches is that the ion path is
obtained by integration, rather
than by the calculation of asymptotes,
and that multiple interactions are,
by the nature of the method, handled in the correct manner.

\section{Rare Event Algorithm}
   A typical dopant profile in a crystalline semiconductor
consists of a near-surface peak followed by an almost exponential
decay over several orders of magnitude in concentration.
If we attempt to directly calculate a statistically significant
dopant concentration at
all depths of the profile we will have
to run many ions that are stopped
near the peak for every one ion that stops in the tail, and most of
the computational effort will not enhance the accuracy of the profile.

   In order to remove this redundancy,
we employ an `atom splitting' scheme\cite{ham64}
to increase the sampling in the deep component
of the concentration profile.
At certain splitting depths in the material, each ion
is replaced by two ions, each with a statistical weighting of
half that prior to splitting.
Each split ion trajectory is run separately, and the weighting of
the ion is recorded along with its final depth.
As the split ions experience
different environments (material is built in front of the ion,
with random
thermal displacements),
the trajectories rapidly diverge from one another.
Due to this scheme, we can maintain the same number of virtual
ions moving at any depth, but their statistical weights
decrease with depth.
During a typical simulation, 1,000 implanted ions are split to yield
around 10,000 virtual ions.

\begin{figure}[!htbp]
\vspace{56mm}
\includegraphics{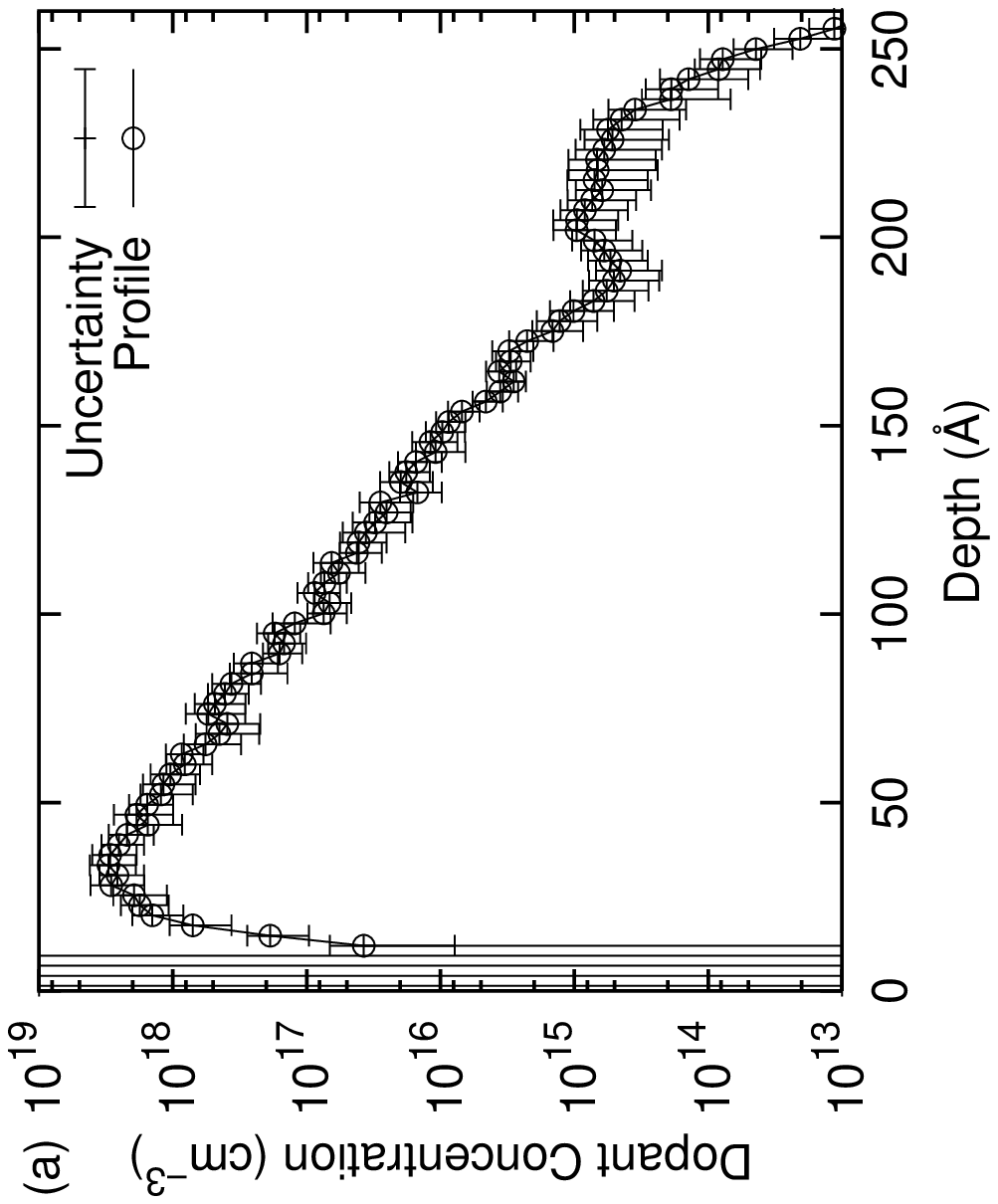}
\includegraphics{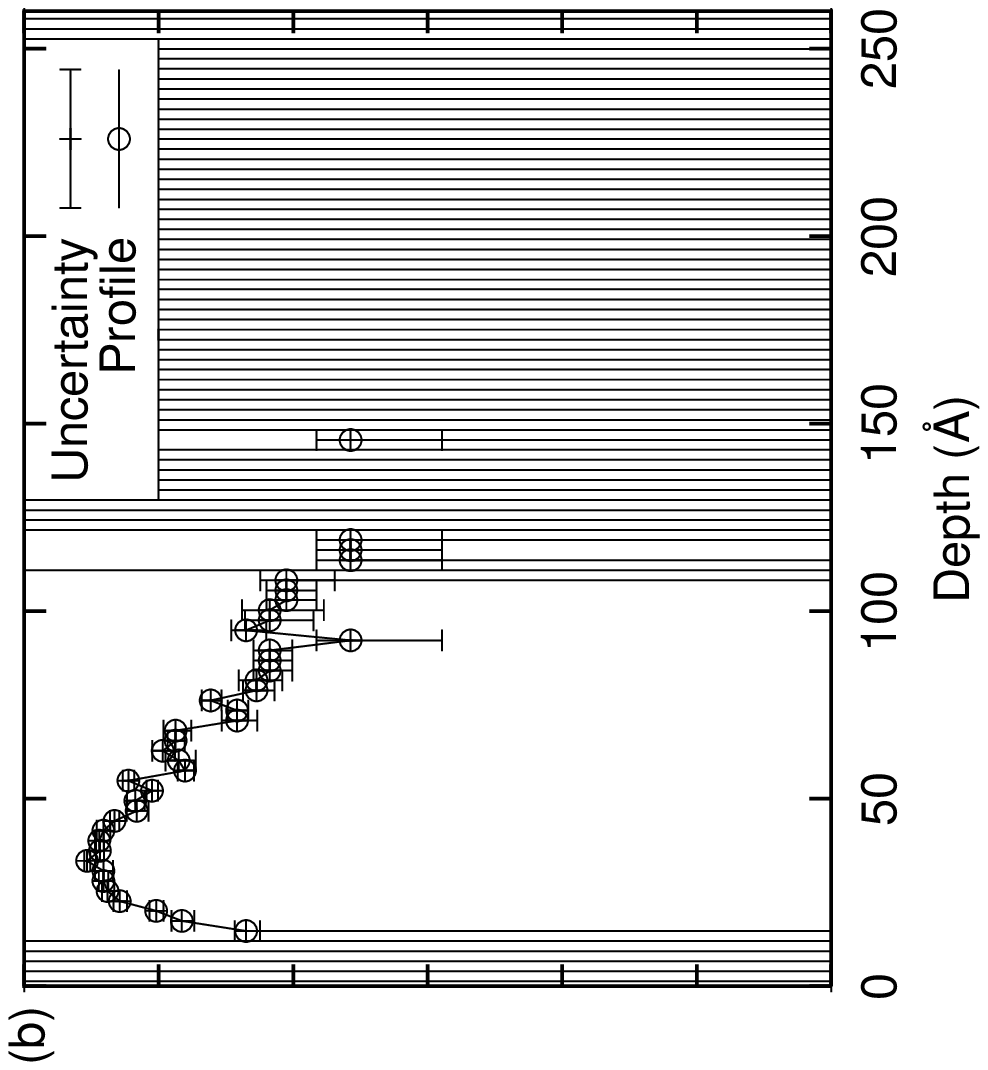}

\caption{
The estimated uncertainty in the calculated dopant profile due to
2 keV As 1$\times$10$^{12}$ cm$^{-2}$ (7,0) implant into Si
with 1 unit cell of surface oxide,
for the same number of initial ions (1,000),
(a) with and (b) without rare event enhancement.}
\label{var}

\end{figure}

We estimate the uncertainty in the calculated dopant profiles
by dividing the final ion depths into 10 sets.
A depth profile is calculated from each set
using a histogram of 100 bins,
and the standard deviation of the
distribution of the 10 concentrations for each bin
is taken as the uncertainty.
Fig.\ \ref{var} demonstrates the effectiveness of the scheme, by
comparing profiles obtained with
and without atom splitting over five orders of magnitude.
We estimate that the rare event algorithm reduces CPU time
by a factor of 90 when calculating profiles over 3 orders of magnitude,
and by a factor of 900 when calculating a
profile over 5 orders of magnitude.

\section{Results and Discussion}

First, we give 2D profiles for low dose, low energy profiles to show
scattering and surface effects.
We then give examples of 1D profiles produced by simulations,
and compare to SIMS data\cite{smsZn,smsSi}.
All simulations were run with a target temperature of 300 K, and
a beam divergence of 1.0$^{\circ}$ was assumed.
Each profile was constructed from 1,000 ions, with the splitting depths
updated every 25 ions,
and a domain of 3$\times$3$\times$3 unit cells was used.
The direction of the incident ion beam is specified by the angle
of tilt, $\theta^{\circ}$, from normal and the
azimuthal angle $\phi^{\circ}$,
as ($\theta$,$\phi$). 
   In the case of the low energy ($\le$ 10 keV) implants, we 
assume one unit cell thickness of surface oxide;
for other cases we assume three unit cells of oxide at
the surface.
For the low energy implants, we have calculated
profiles over a change of five orders of magnitude in concentration; for
the higher energy implants we calculate
profiles over 3 orders of magnitude.
A dose of 1$\times$10$^{12}$ cm$^{-2}$ (zero damage) is used unless
otherwise noted.

\begin{figure}[!htbp]
\vspace{55mm}
\includegraphics{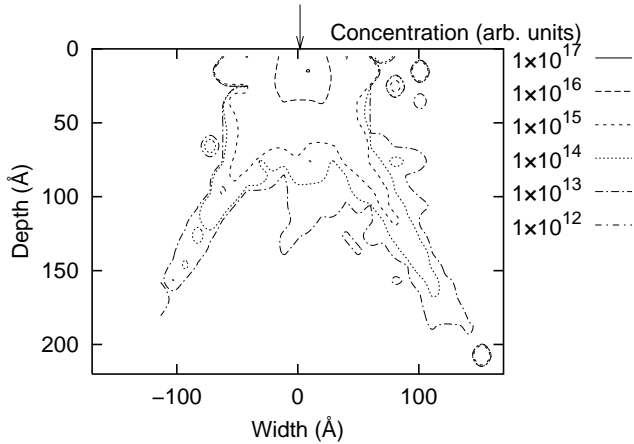}

\caption{
Calculated 2D dopant profile due to
0.2 keV B (0,0) implant into Si.}
\label{B0t0r200}

\end{figure}

\begin{figure}[!htbp]
\vspace{58mm}
\includegraphics{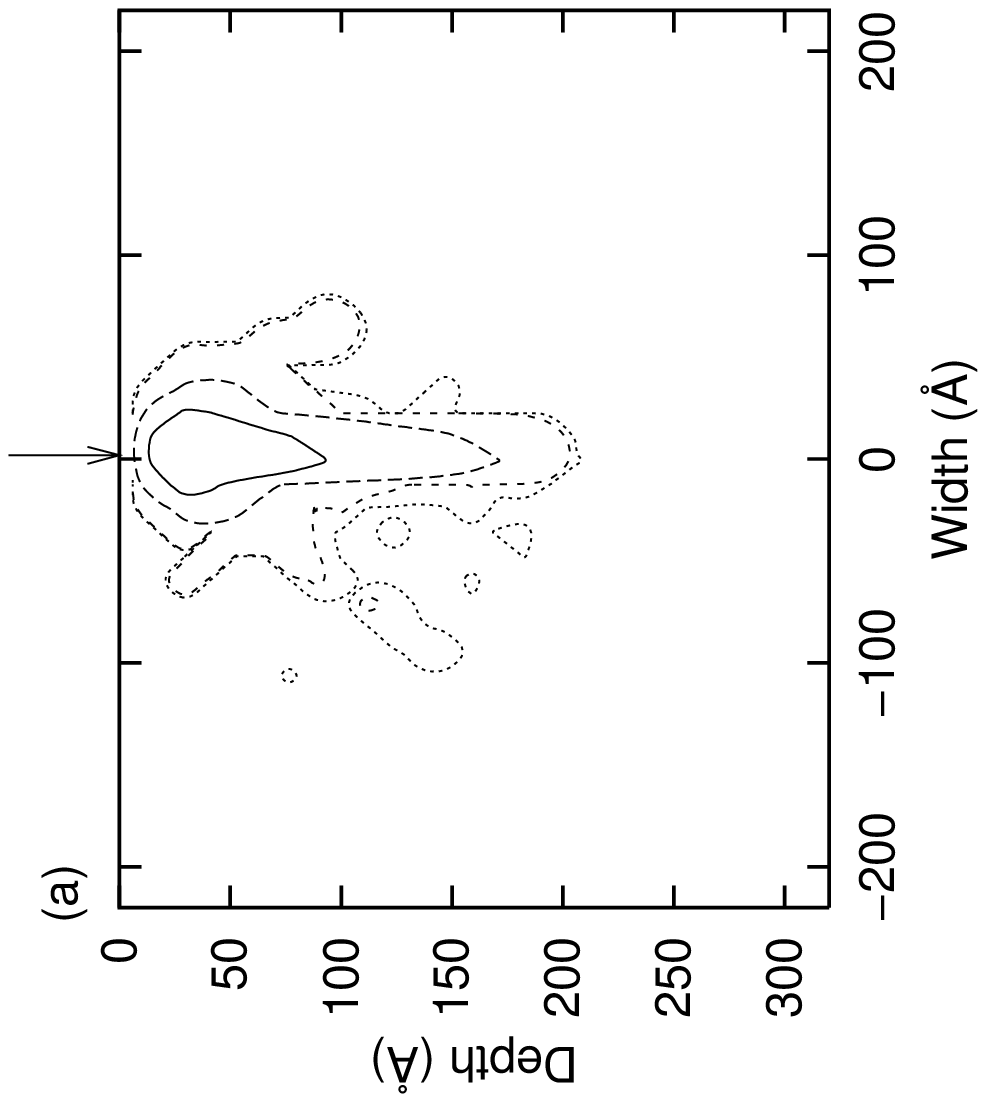}
\includegraphics{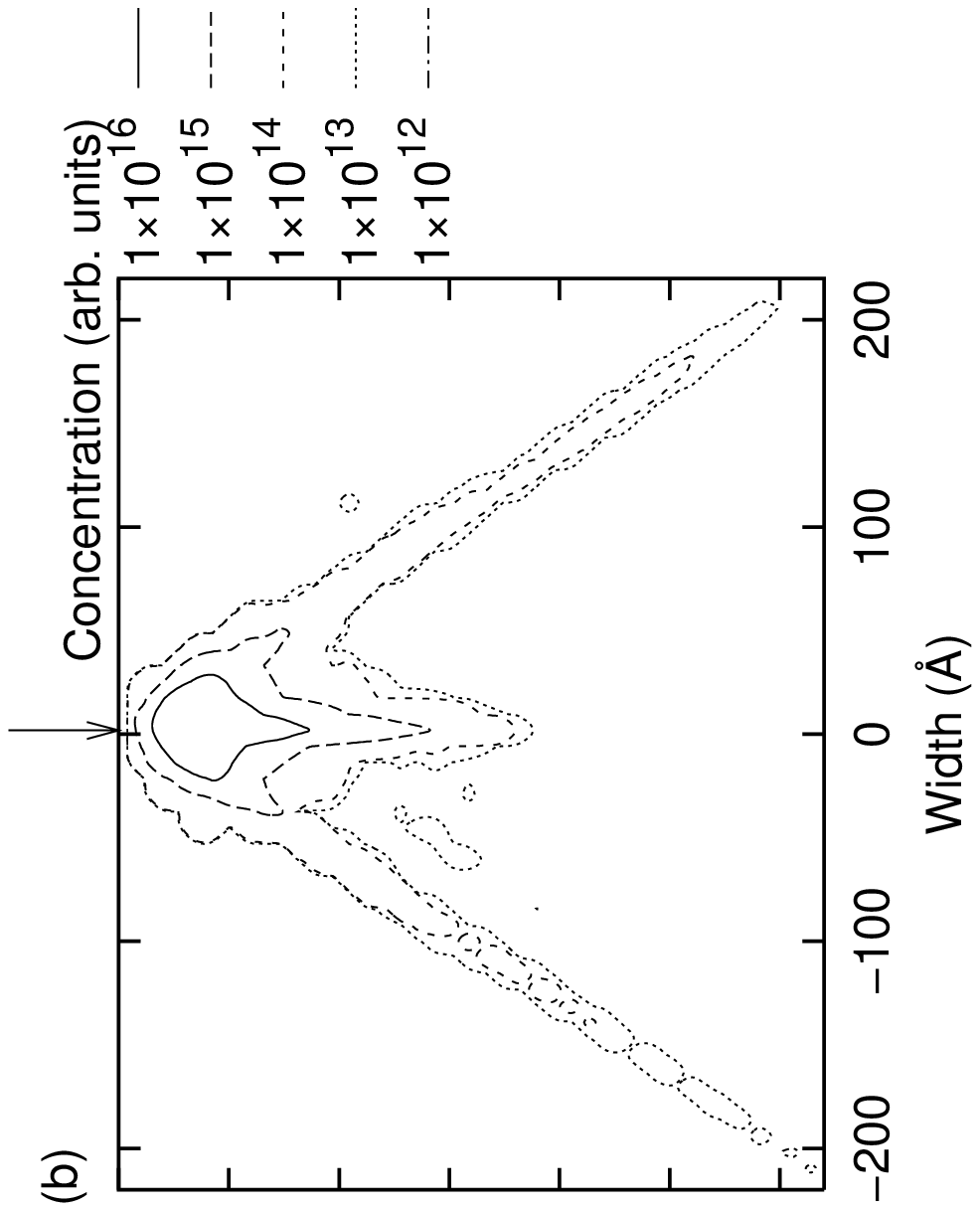}

\caption{
Calculated 2D dopant profiles due to
2 keV As (0,0) implants into Si,
with (a) 1 and (b) 3 unit cells of surface oxide.}
\label{A0t0r2K}

\end{figure}

\begin{figure}[!htbp]
\vspace{55mm}
\includegraphics{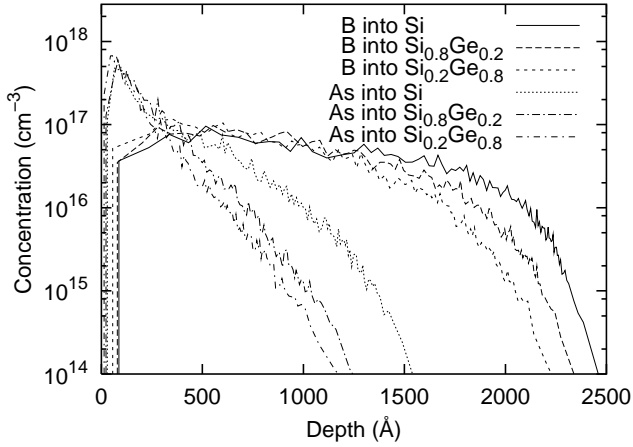}

\caption{
Calculated dopant profiles for 10 keV B and As
(0,0) implants into Si$_{1-x}$Ge$_x$.}
\label{BAs10KSiGe}

\end{figure}

   2D profiles are shown projected onto
the plane normal to the surface and
containing the zero degree azimuth. This makes it easy to differentiate
between major channeling directions; the $\langle$100$\rangle$ channel
is vertical, and the four $\langle$110$\rangle$ channels
appear at angles of 35$^{\circ}$ from vertical.
Fig.\ \ref{B0t0r200} shows the result of an
ultra-low energy implant into Si.
Although the implant is in the
$\langle$100$\rangle$ direction, this channel
is closed at such low ion energies and the only channeling occurs in the
$\langle$110$\rangle$ direction. This demonstrates the need to have a
`universal' electronic stopping model, rather than a model tuned for a
particular channeling direction.
The effect of the amount of surface
disorder is shown in Fig.\ \ref{A0t0r2K}.
Increasing the thickness of the surface layer leads to more ions being
scattered into the larger $\langle$110$\rangle$ channel, and hence gives
a far deeper tail to the profile.

\begin{figure}[!htbp]
\vspace{55mm}
\includegraphics{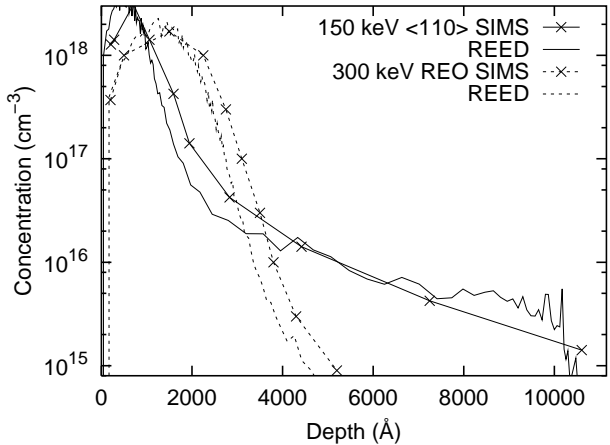}

\caption{
Calculated and experimental\protect\cite{smsZn} dopant profiles due to
Zn 3$\times$10$^{13}$ cm$^{-2}$ implants into GaAs.}
\label{Zn-GaAs}

\end{figure}

\begin{figure}[!htbp]
\vspace{55mm}
\includegraphics{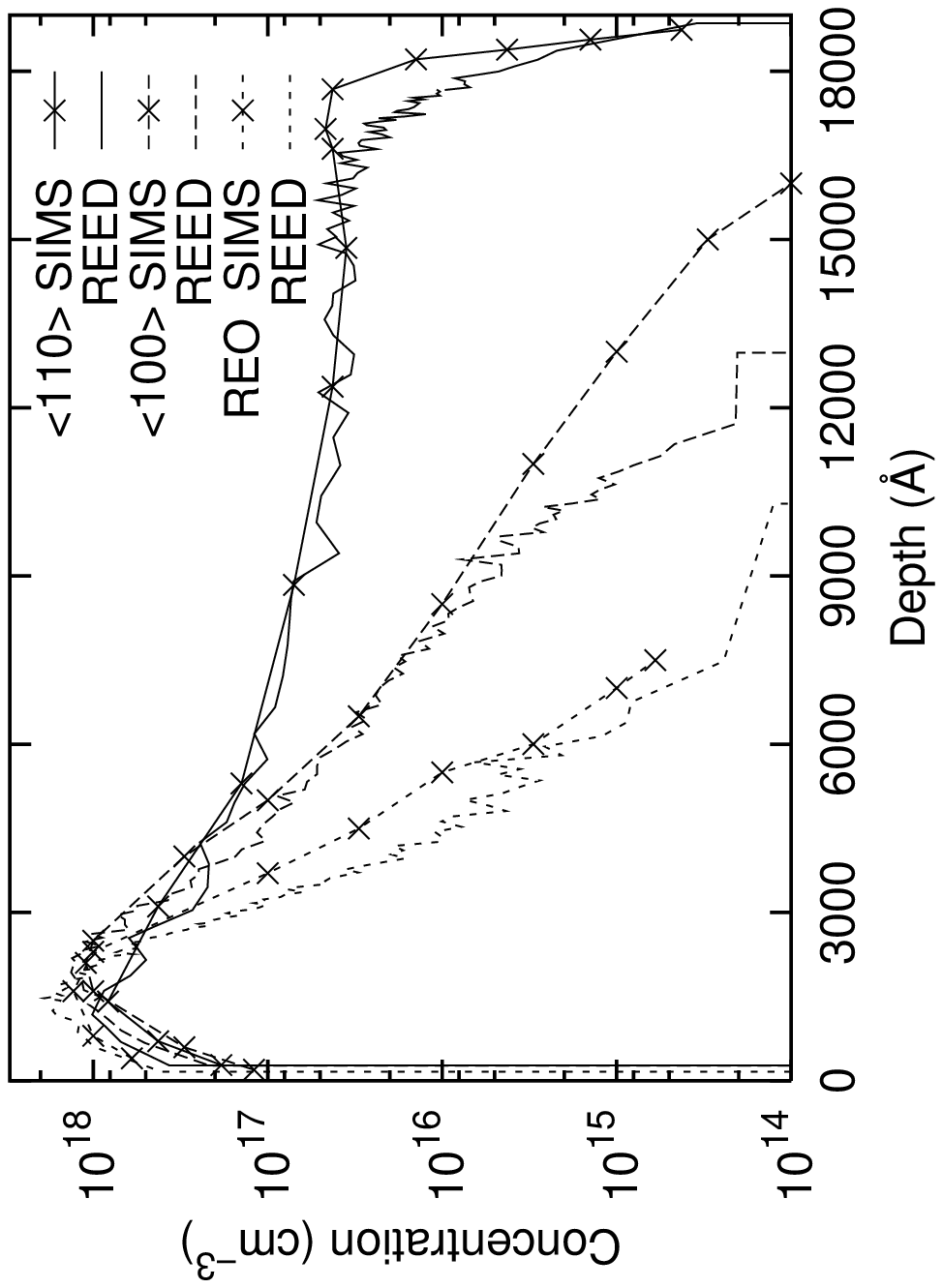}

\caption{
The calculated and experimental\protect\cite{smsSi}
dopant profiles due to
150 keV Si 3$\times$10$^{13}$ cm$^{-2}$ implants into GaAs.}
\label{Si-GaAs}

\end{figure}

%
%

\begin{figure}[!htbp]
\vspace{55mm}
%
\includegraphics{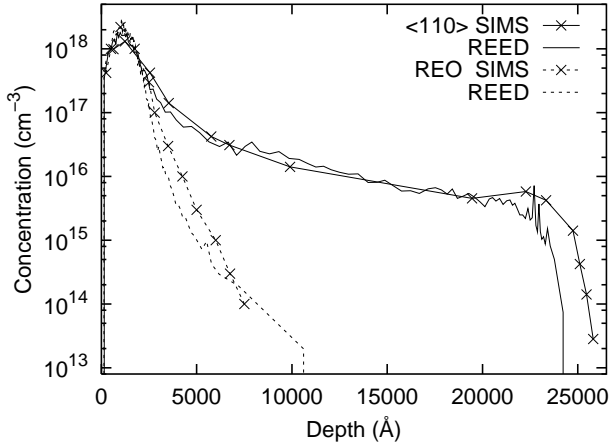}

\caption{
The calculated and experimental\protect\cite{smsSi}
dopant profiles due to
300 keV Se 3$\times$10$^{13}$ cm$^{-2}$ implants into GaAs.}
\label{Se-GaAs}

\end{figure}

There is increasing interest in the use of SiGe as a replacement
for Si currently used in
CMOS technology, due to its higher switching speed\cite{hol97}.
Fig.\ \ref{BAs10KSiGe} shows the effect of Ge concentration on profiles
from B and As implants into Si$_{1-x}$Ge$_x$ targets.
The trend is clearly for shallower profiles with increasing Ge
concentration, but this is extremely non-linear; the difference between
$x=0$ and $x=0.2$ profiles is greater than the difference between
$x=0.2$ and $x=0.8$ profiles.
The remaining figures show the calculated concentration profiles of
several ion species implanted under various conditions into
GaAs substrates, and comparison with available SIMS data.
The results of the REED calculations show good agreement with the
experimental data, demonstrating the accuracy of our model and its
transferability to many ion-target combinations and implant conditions.

\section{Conclusions}
  In summary,
we have developed a restricted MD code to simulate the
ion implant process and calculate `as implanted' dopant profiles.
This gives us the
accuracy obtained by time integrating atom paths, with an
efficiency far in excess of full MD simulation.

The scheme described here gives a viable
alternative to the BCA approach.
Although it is still more expensive computationally, it is sufficiently
fast to be used on modern desktop computer workstations.
The method has two major advantages over the BCA approach:
(i) Our MD model consists only of standard empirical
potentials developed for bulk semiconductors
and for ion-solid interactions.
The only fitting is in the electronic stopping model, and
this involves \emph{only one} parameter per ion-target combination.
   We believe that by using physically based models for all aspects
of the program, with the minimum of fitting parameters, we obtain good
transferability to the modeling of implants outside of our fitting set.
(ii) The method does not break down at the
low ion energies necessary for
production of the next generation of
semiconductor technology; it gives the
correct description of multiple, soft interactions that occur both
in low energy implants, and higher energy channeling.
Hence our method remains truly predictive at these low ion energies, whilst
the accuracy of the BCA is in doubt.

\begin{ack}
This work was performed under the auspices of the
United States Department of Energy.
\end{ack}

\end{document}